\documentclass[runningheads]{llncs}

\usepackage{amssymb,amsmath,amsfonts}
\usepackage{graphicx}
\usepackage{url}
\usepackage{color}
\usepackage{amsfonts}
\usepackage{blindtext}
\usepackage{mathrsfs}
\usepackage{bbm}
\usepackage{url}
\usepackage{paralist}
\usepackage[english]{babel}
\usepackage{booktabs}
\usepackage{multicol}
\usepackage{multirow}
\usepackage{subfig}
\begin{document}
\title{Siamese Neural Networks for Class Activity Detection}
%
%
\author{Hang Li\inst{1} \and
Zhiwei Wang\thanks{Work was done when the authors did internship in TAL Education Group}\inst{2} \and
Jiliang Tang\inst{2} \and Wenbiao Ding\inst{1} \and Zitao Liu\thanks{Corresponding Author: Zitao Liu}\inst{1}}

\authorrunning{H. Li et al.}
%

\institute{TAL Education Group, Beijing, China  \\
\email{\{lihang4,dingwenbiao,liuzitao\}@100tal.com}
\and Data Science and Engineering Lab, Michigan State University, USA\\
\email{\{wangzh65,tangjili\}@msu.edu}}
\maketitle              
\begin{abstract}

Classroom activity detection (CAD) aims at accurately recognizing speaker roles (either teacher or student) in classrooms. A CAD solution helps teachers get instant feedback on their pedagogical instructions. However, CAD is very challenging because (1) classroom conversations contain many conversational turn-taking overlaps between teachers and students; (2) the CAD model needs to be generalized well enough for different teachers and students; and (3) classroom recordings may be very noisy and low-quality. In this work, we address the above challenges by building a Siamese neural framework to automatically identify teacher and student utterances from classroom recordings. The proposed model is evaluated on real-world educational datasets. The results demonstrate that (1) our approach is superior on the prediction tasks for both online and offline classroom environments; and (2) our framework exhibits robustness and generalization ability on new teachers (i.e., teachers never appear in training data).

\keywords{Multimodal learning \and Neural networks \and Class activity detection.}
\end{abstract}

\section{Introduction}
\label{sec:intro}

It is essential to equip instructor training with informative dialogic feedback on their classroom activities, which allows teachers to adjust and refine their teaching instructions \cite{scheeler2004providing,freiberg1988alternative,akalin2015effects,10.1145/3366423.3380018,chen2019multimodal}. Prior researches have been demonstrated that pedagogical teaching styles and instructions may significantly influence students' engagements and academic achievements \cite{tanner2013structure,lockheed1987school,owens2017classroom}. Traditionally, providing such feedback is very logistically complex and expensive, as it heavily relies on human annotations \cite{brinko1993practice,kane2012gathering,nystrand2006research,nystrand}. This makes it inapplicable in  real-world education scenarios. Thus, in this work, we focus on building an automatic AI driven solution to solve this fundamental class activity detection (CAD) problem. More specifically, we aim at automatically annotating classroom audio recordings by recognizing different speakers' roles, i.e., student or teacher. CAD solutions produce basic information about the quantities and distributions of classroom conversations, which are one of the essential steps for deep classroom analysis \cite{li2020multimodal}.

A large spectrum of models have been developed to solving the CAD problem \cite{owens2017classroom,cosbey2019deep,donnelly2016automatic,bergman2015comparing}. Owens et al. proposed a machine learning algorithm that captures distinctive patterns in different instructional techniques and classifies the classroom sound into different class activities \cite{owens2017classroom}. Cosbey et al. targeted on the same classroom sound classification problem as in \cite{owens2017classroom} and adopted deep recurrent neural networks to extract meaningful features from audio frames \cite{cosbey2019deep}. Wang et. al conducted CAD by using LENA system \cite{ganek2016language} and identified three discourse activities of teacher lecturing, class discussion and student group work \cite{wang2014automatic}.

However, CAD in real-world scenarios is still extremely difficult because of three challenges: (1) \textit{conversational turn-taking overlap}: Classroom conversations usually contain many frequent talk exchanges between teachers and students, which leads to a number of inextricable speech overlaps; (2) \textit{vocal variability and uniqueness}: Every person's voice is different and unique, which poses a difficult question on the generalization ability of the CAD solution; and (3) \textit{classroom noise}: Both online and offline classrooms in reality are dynamic, complex and noisy. In the attempt to solve the aforementioned challenges, we develop the Siamese neural framework to precisely detect teacher and student activities from classroom audio recordings. The contributions of this work are summarized as follows: (1) It presents a pioneer research on the CAD problem and proposes a novel Siamese neural framework to tackle this problem; and (2) we comprehensively evaluate our framework with different realizations and their benefits on both online and offline real-world, large-scale classroom datasets.

\vspace{-0.1cm}
\section{The Siamese Neural Framework}
\vspace{-0.1cm}
\label{sec:method}

\begin{figure}
  \centering
  \includegraphics[width=\textwidth]{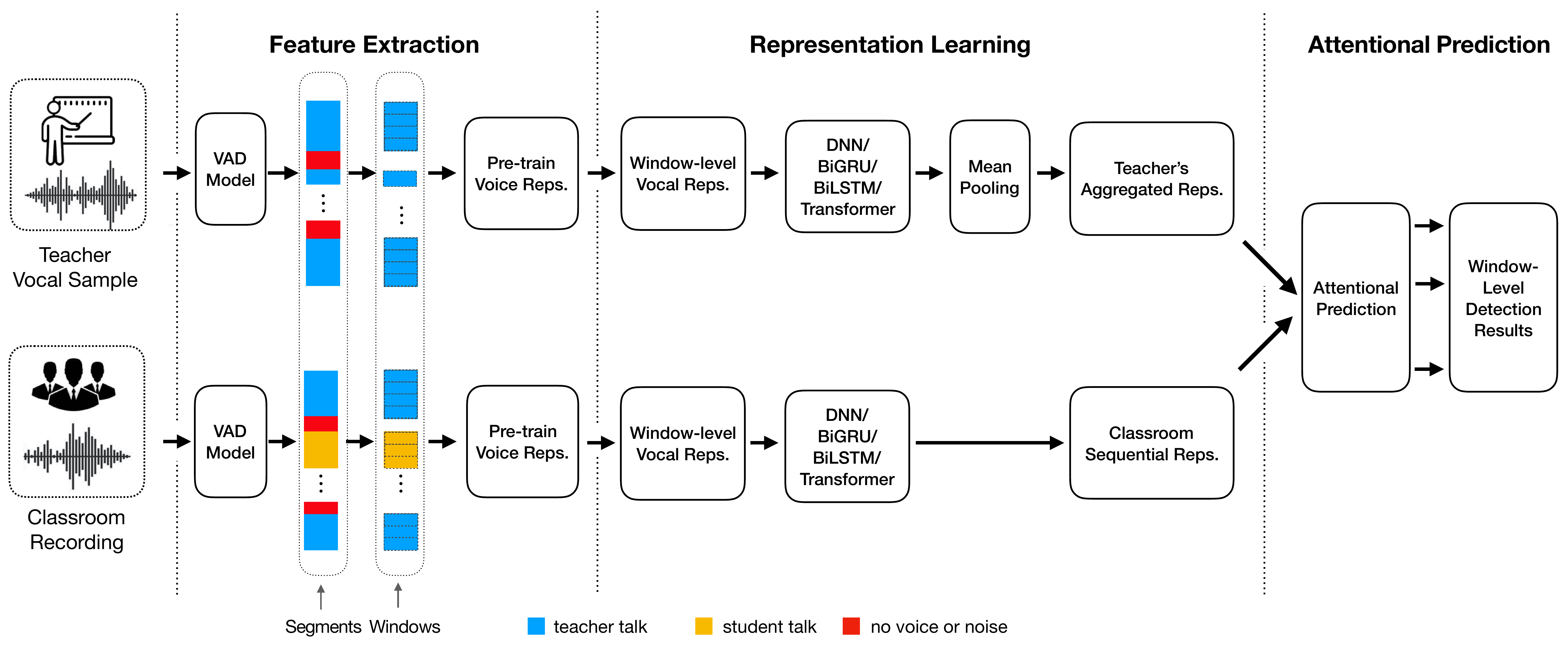}
  \caption{The overview of our Siamese neural framework. VAD is short for voice activity detection.}
  \label{fig:framework}
  \vspace{-0.6cm}
\end{figure}

In this section, we describe our end-to-end Siamese neural framework for the CAD problem in details. Our framework consists of three key components: (1) feature extraction module that extracts window-level raw embeddings from a pre-train large-scale audio encoding neural network; (2) the representation learning module, which extracts semantic representations from each classroom audio segment; and (3) an attentional prediction module that predicts the activity type for each window. The overall framework architecture is shown in Figure \ref{fig:framework}.

\noindent{\bf Feature Extraction} We first utilize a well-studied voice activity detection (VAD) system to segment audio streams into pieces of utterances and filter out the noisy and silent ones \cite{sohn1999statistical,tanyer2000voice,ramirez2004efficient}. Then we transform each segment into frames of pre-defined width and step, and log-mel-filterbank energies of dimension 40 are extracted from each frame. After that, we obtain windows by using non-overlapping sliding windows of a fixed length on these frames. Once we create these audio windows from both teachers' vocal sample segments and classroom recording segments, we extract windows' corresponding low-dimensional dense vocal representations from a pre-trained acoustic neural network.

\noindent{\bf Representation Learning} We learn a refined vocal representation for each window by utilizing the contextual dependencies within each segment (either from teachers' vocal samples or classroom recordings). In our framework, any existing sequential modeling function such as long short-term memory (LSTM), gated recurrent unit (GRU), etc. can be used \cite{devlin2018bert,hu2018squeeze,velivckovic2017graph}. By considering the contextual windows across entire segment, we are able to model the changes of tones and pitches in the audio stream smoothly and reduce the noises and outliers in the raw feature extraction component.

\noindent{\bf Attentional Prediction} We design an attentional prediction module focusing on the window-level class activity detection tasks. Our attentional prediction module is inspired by the intuition that all the audio windows spoken by the teacher share common attributes that are very different from those shared from student's audio windows. Thus, we use teachers' vocal samples as an aggregated query and compute an attention score with each individual window from classroom recordings. The higher the attentional score is, the more likely the audio window is spoken by the teacher. Based on this idea, we first add a mean pooling layer to aggregate all the teacher's vocal sample representations. This yields a robust and representative query embedding of the teacher's voice signals. The obtained vector is used as a voice biometrics query to compute attention scores with each individual window representation. In order to effectively train our framework, we design a cross-entropy loss function as the optimization objective. We use mini-batch stochastic gradient decent algorithm to minimize the objective and update the our model parameters.

\section{Experiments}
\label{sec:experiment}

We evaluate our framework with two real-world K-12 education datasets: (1) the \emph{online} dataset, which includes 400 classroom recordings and 300 distinct teachers from a third-party online education platform\footnote{https://www.xes1v1.com/}; and (2) the \emph{offline} dataset that includes 100 recordings and 36 distinct teachers from physical offline classrooms. We randomly select 100 and 10 recordings from \emph{online} and \emph{offline} dataset respectively as our test sets. The prediction results are denoted as ``Main''. Moreover, in order to evaluate the model generalization ability to new teachers, we further filter out teachers from above test set if the teachers appear in the training set and the prediction results are denoted as ``Generalization''. We choose to use area under curve (AUC) score to evaluate the model performance \cite{fawcett2006introduction}. 

We choose the following approaches as our baselines: (1) \emph{Average}: Vocal representations from feature extraction component are directly used for attentional prediction; (2) \emph{DNN/GRU/LSTM}: A single layer fully connected neural network/a bidirectional GRU/a bidirectional LSTM is used in the representation learning component \cite{murtagh1991multilayer,chung2014empirical,hochreiter1997long}. We use 128 neurons and ReLU as the activation function; and (3) \emph{Transformer}: A transformer is used in the representation learning component \cite{vaswani2017attention}. We choose to use 2 layers in the transformer and set 4 heads for each layer. We set the dimension of each head to 16.

\noindent{\bf Experimental Results}: The results are shown in Table \ref{tab:overl_result}. For the main task, we find that (1) the \emph{Average} performs much worse than any other method. This suggests that the fine-tuned representation learning plays an important role in the final prediction; (2) compared to \emph{GRU}, \emph{LSTM}, and \emph{Transformer}, \emph{DNN} has achieve a lower detection accuracy. This is expected as it is not able to capture the contextual information of windows within each segment; (3) the performance of all methods on \emph{online} dataset is generally better than results on \emph{offline} dataset. We argue that this is because the signal to noise ratio of offline recordings is much higher than the ratio in online recordings \cite{li2020multimodal}; and (4) both \emph{GRU} and \emph{Transformer} have comparable performance, which is consistent with the previous findings \cite{Karita2019Comparative}. For the generalization task, we have similar observations. The high accuracy achieved by Transformer and LSTM demonstrates the generalization ability of the proposed framework.

\vspace{-0.9cm}
\begin{table}[!hbpt]
\small
    \caption{Experimental results on the \emph{online} and \emph{offline} datasets.}
    \vspace{-0.4cm}
    \label{tab:overl_result}
    \begin{center}  
    \begin{tabular}{ccccccc}
    \toprule
    Task & Dataset & Average  & DNN & GRU &  LSTM &Transformer\\ 
    \midrule                  
   \multirow{2}{*}{Main} & online     &0.895  &0.926  &0.936  &0.933  & 0.942 \\                    
   & offline    & 0.713 & 0.810 & 0.881   & 0.858  & 0.858 \\
    \midrule                  
    \multirow{2}{*}{Generalization} &online  &0.895  &0.922  &0.932  &0.931   & 0.937 \\                  
     & offline    & 0.749 & 0.840 & 0.880   & 0.805  & 0.882 \\
    \bottomrule
    \end{tabular}
    \end{center}
\end{table}

\vspace{-1.3cm}

\section{Conclusion}
\label{sec:conclusion}
\vspace{-0.1cm}
We present a Siamese framework to tackle the CAD problem. Experiments demonstrate both detection performance and generalization ability of our framework. In the future, we would like to design models that can combine both audio and video data to generate more comprehensive classroom activity feedback.

\vspace{-0.3cm}
\section*{Acknowledgements}
Zhiwei Wang and Jiliang Tang are supported by the National Science Foundation of United States under IIS1714741, IIS1715940, IIS1715940, IIS1845081 and IIS1907704. 

%
%
%
%
\bibliographystyle{splncs04.bst}
\bibliography{aied2020}
\end{document}